\documentclass[useAMS,usenatbib]{mn2e}
\usepackage{ctable}
\usepackage{float}
\usepackage{graphicx}
\usepackage[caption=true]{subfig} 
\usepackage{color}
\usepackage{amsmath}
\usepackage{ctable}
\newcommand{\kms}{\ensuremath{\mathrm{km\,s^{-1}}}}

\newcommand{\sigmastar}{\sigma_{\star}}
\newcommand{\Mcutoff}{$M_{\mathrm{low} }$} 
\newcommand{\Msun}{M_\odot}

\newcommand{\MoLs}{\Upsilon_{\star}}

\newcommand{\Rein}{$R_{\rm Ein }$}
\newcommand{\fdm}{$f_{\rm DM }$}

\newcommand{\Reff}{$R_{\rm eff }$}

\title[XLENS II. Sample selection and kinematics]{The X-shooter Lens Survey - II. Sample 
presentation and spatially resolved kinematics\thanks{Based on observations collected 
at the European Organisation for Astronomical Research in the Southern Hemisphere, 
Chile. (P086.A-0312, PI: Koopmans  and P089.A-0364, PI: Spiniello)}}
\author[C.Spiniello et al.]{C. Spiniello$^{1}$\thanks{E-mail:spini@mpa-garching.mpg.de}, 
L.V.E. Koopmans$^{2}$, S.C. Trager$^{2}$, M.Barnab{\`e}$^{3,4}$, T. Treu$^{5,6}$,
\\
\\
{\LARGE \normalfont O. Czoske$^{7}$, S.Vegetti$^{1}$, A.Bolton$^{8}$ }
\\
$^{1}$Max-Planck Institute for Astrophysics, Karl-Schwarzschild-Strasse 1, 85740 Garching, Germany\\
$^{2}$Kapteyn Astronomical Institute, University of Groningen, P.O. Box 800, 9700 AV Groningen, the Netherlands\\
$^{3}$Dark Cosmology Centre, Niels Bohr Institute, University of Copenhagen, Juliane Maries Vej 30, 2100 Copenhagen {\O}, Denmark \\
$^{4}$Niels Bohr International Academy, Niels Bohr Institute, University of Copenhagen, Blegdamsvej 17, 2100 Copenhagen {\O}, Denmark \\
$^{5}$Department of Physics, University of California Santa Barbara, Santa Barbara,CA 93106, USA\\ 
$^{6}$Department of Physics and Astronomy, University of California, Los Angeles, CA 90025-1547, USA\\
$^{7}$Institut  f\"{u}r Astrophysik, Universit\"{a}t Wien, T\"{u}̈rkenschanzstra$\beta$e 17, 1180 Wien, Austria \\
$^{8}$Department Physics and Astronomy, University of Utah, UT 84112, USA\\}
\begin{document}

\date{Accepted Year Month Day. Received Year Month Day; in original form Year Month Day}
\pagerange{\pageref{firstpage}--\pageref{lastpage}} \pubyear{2015}
\maketitle
\label{firstpage}

\begin{abstract}
We present the X-shooter Lens Survey (XLENS) data. The main goal of XLENS is to disentangle 
the stellar and dark matter content of massive early-type galaxies (ETGs), through combined strong gravitational lensing, 
dynamics and spectroscopic stellar population studies. 
The sample consists of 11 lens galaxies covering the redshift 
range from $0.1$ to $0.45$ and having stellar velocity dispersions between $250$
and $380\,\mathrm{km}\,\mathrm{s}^{-1}$. 
All galaxies have multi-band, high-quality HST imaging. 
We have obtained long-slit spectra of the lens galaxies 
with X-shooter on the VLT.  
We are able to disentangle the dark and luminous mass components by combining lensing and extended kinematics data-sets, 
and  we are also able to precisely constrain stellar mass-to-light ratios and infer the value of the low-mass cut-off of the IMF, 
by adding spectroscopic stellar population information. 
Our goal is to correlate these IMF parameters with ETG masses and investigate the relation between baryonic and 
non-baryonic matter during the mass assembly and structure formation processes. 
In this paper we provide an overview of the survey, highlighting its scientific motivations, main goals 
and techniques. We present the current sample, briefly describing the data reduction and analysis process, 
and we present the first results on spatially resolved kinematics. 

\end{abstract}


\section{Introduction}
Massive early-type galaxies (ETGs, $\sigma \geq 200\,\mathrm{km}\,\mathrm{s}^{-1}$) 
contain more than half of the stellar mass in the Universe (e.g., \citealt{Bell2003}). 
Their internal structure and dynamics as well as the relationship between 
baryonic luminous matter and dark matter therefore provide key quantitative tests 
of the $\Lambda$CDM cosmological paradigm (\citealt{Blumenthal1984}) and are 
crucial to fully comprehend the processes involved in the hierarchical formation of 
galaxies (e.g.\ \citealt{White1978, Davis1985, Frenk1985}). \\
\indent In hierarchical structure formation models massive galaxies are believed to form through 
mergers of lower-mass galaxies (e.g.,\ \citealt{Blumenthal1984}), however their stellar populations 
are old and evolved (see \citealt{Renzini2006} and \citealt{Conroy2013} for recent reviews). 
In addition, analysis of the evolution of the stellar mass and the luminosity functions 
has demonstrated that there is little evolution in the co-moving space density of massive passive 
galaxies since $z\sim1$  (\citealt{Cimatti2004,McCarthy2004, Glazebrook2004, Daddi2005,
Saracco2005,  Bundy2007,PerezGonzalez2008,Marchesini2009,Ferreras2009,Carollo2013,Muzzin2013,Lundgren2014}).   
The prevailing idea is that ETGs formed the bulk of their stars early in the evolution of the Universe 
(i.e., cluster ETGs $z>3$, field ETGs at $z>1.5-2$, \citealt{Thomas2005, Renzini2006,Cimatti2008,Whitaker2013}) 
 and then  only at lower redshift  (i.e., $z<1$) may have merged together to build the massive 
ETGs that we see today in the nearby Universe (\citealt{Johansson2012,Nipoti2012,Buitrago2013}). 

To reconcile these apparently conflicting results and to paint a more robust physical picture for the 
formation and evolution of these massive systems, enormous effort has been expended in the last 
two decades to constrain the relative contributions of baryonic, dark matter and black hole constituents 
of ETGs through stellar dynamical tracers, X-ray studies, and
gravitational lensing (e.g.\ \citealt{Fabbiano1989, Mould1990, Saglia1992, Franx1994, Carollo1995, 
Arnaboldi1996,Rix1997, Loewenstein1999, Gerhard2001,Romanowsky2003,  Treu2004, Treu2006,  Cappellari2006, Thomas2007, 
Czoske2008, Auger2010, Auger2010b, Coccato2010,Barnabe2011,Das2011,Treu2010b, Courteau2013}). 
A variety of recent observations suggest that baryonic luminous matter, 
which dominates astrophysical observables, and dark matter, which dominates 
the dynamics during galaxy formation, are strongly linked.
As a consequence, progress in understanding ETGs and more generally 
the theory of galaxy formation and evolution requires the precise measurement 
of their stellar and dark matter density profiles, 
together with measurements of how these quantities depend upon properties 
such as the mass and the age of the stellar assembly. 

Lensing and dynamics observations show that in the inner region 
 of massive ETGs (inside few effective radii, \Reff), where baryonic and dark matter are both present, 
the total mass density profile is well described by a power-law density profile, 
$\rho_{\rm tot} \propto r^{-\gamma'}$, with an almost isothermal slope of $\gamma' \approx 2$ and $\sim10\%$
intrinsic scatter (e.g.\ \citealt{Gerhard2001,Treu2004, Koopmans2006, Koopmans2009, Coccato2009, Auger2010,  
Barnabe2009, Barnabe2011, Napolitano2011,Bolton2012, Sonnenfeld2012,DuttonTreu2014}). 
However, in the same region, the dark matter density profile seems to be 
shallower, consistent with a density slope $1.0 < \gamma_{\rm DM} < 1.7$,  although this slope is less well constrained 
than the total mass density slope (e.g.\ \citealt{ Treu2004,Dye2005,Grillo2012,Sonnenfeld2012, Sonnenfeld2014, Sonnenfeld2015}).

In addition, recent observations as well as theoretical studies based on stellar population and dynamical models 
(e.g.\ \citealt{Bullock2001, Padmanabhan2004}) indicate that, {\sl for a universal IMF}, 
the dark matter fraction (\fdm) in the internal region increases monotonically with the mass of the galaxy 
(e.g.\ \citealt{Zaritsky2006, Auger2010b}), a trend that is more conspicuous in the case of slow-rotator ellipticals
(\citealt{Tortora2009}).  
On the other hand, lensing and dynamics studies also suggest that the luminous stellar 
mass-to-light ratio ($\MoLs$) scales with the luminous mass of the  system (\citealt{Davis1985,Bardeen1986,  Bell2001,Girardi2002,
Napolitano2005, Grillo2009, Auger2010}), under the assumption of standard  Navarro-Frenk-White dark matter halo profile (NFW, \citealt{Navarro1996}).

Despite this progress, it remains difficult to separate the stellar and
dark matter components, mostly due to a relatively poor understanding of the precise shape of the stellar IMF and its
associated $\MoLs$. Uncertainties related to the latter can easily lead to a factor of  two uncertainty on the
inferred stellar mass.

Although it is commonly assumed that the IMF 
is universal and independent of cosmic time (e.g., \citealt{Kroupa2001, Chabrier2003,Bastian2010}), several authors 
have suggested that the  IMF might  evolve (\citealt{Dave2008,  vandokkum2008}) or depend on 
the stellar mass of the system (e.g.\ \citealt{Worthey1992, Trager2000b,Treu2010, Graves2009, Graves2010, Auger2010b, Napolitano2010, Cappellari2012,Martin2015,Posacki2015}).
Recent results based on single stellar population (SSP) modelling of galaxy spectra, 
indicate that the number of low-mass stars ($M < 0.3\,M_{\odot}$) increases more rapidly 
with galaxy mass than the number of high-mass stars (\citealt{vandokkum2010}, hereafter vDC10, 
\citealt{Spiniello2011, Spiniello2012, Spiniello2014a, Cappellari2012, Tortora2013, Ferreras2013, LaBarbera2013}). 
This could imply that part of the increase in $\MoLs$ with
galaxy mass can be due to a changing stellar IMF rather than only an increasing dark matter fraction, 
consistent with the earlier findings by \citet{Treu2010b} and Auger et al.$\,$(2010b), based on combing lensing, 
dynamics and stellar population information. 
The shape of the IMF and the internal fraction of dark matter mass inside massive ETGs can not be both universal at the same time, 
but it remains difficult to break the degeneracy between these two effects.

The X-shooter Lens Survey (XLENS, \citealt{Spiniello2011, Spiniello2012}) aims to take these analyses one step further.
With a combined analysis of strong gravitational lensing, dynamical and spectroscopic stellar population studies, 
the XLENS project will be able to disentangle the stellar and dark matter content of galaxies 
and, for the first time, directly constrain the normalization, shape and cut-off mass (\Mcutoff) 
of the low-mass end of the IMF and correlate these results with other galaxy properties such as 
galaxy mass, size, stellar density and/or stellar velocity dispersion 
(e.g., \citealt{Spiniello2011, Spiniello2012, Spiniello2014a, Barnabe2013, Spiniello2015}).

In particular, our methodology consists of combining spatially resolved kinematics 
with high-precision strong-gravitational-lensing measurements of the total mass 
and the mass density profile near the lens Einstein radius (\Rein), 
to obtain a precise internal dark-matter mass fraction inside one effective radius. 
By taking advantage of the wide wavelength coverage and throughput of the X-shooter spectrograph, 
we are able to obtain high-resolution spectra from the UVB to the near-IR for detailed stellar population and 
spatially-resolved kinematics studies of the high-mass end of the early-type galaxies (ETGs) up to redshift $z\sim 0.7$. 

A previously under-appreciated result of our approach is that it allows us to tackle 
the long-standing problem of constraining the low-mass cut-off (\Mcutoff) for the IMF. 
In previous studies, the value of \Mcutoff\, has always been treated as a relatively unconstrained parameter, despite 
being critical to determine the value of $\MoLs$. 
Stars with masses below $\sim 0.15 \Msun$ have very little effect on the spectral lines and on the line-index measurements
in the optical and near-infrared for any assumed IMF slope (\citealt{Conroy2012}), 
but they provide a substantial contribution to the total mass budget of the system for IMF slopes steeper than $\sim 2$ (\citealt{Worthey1994}). 
Therefore, to constrain the value of \Mcutoff\, it is necessary to support spectroscopic studies with 
a combined lensing and dynamics analysis (see \citealt{Barnabe2013}). 

In this paper we provide an overview of the survey, focusing on the data reduction and analysis process. 
We present scientific results on the IMF slope and cut-off mass of the full sample in \citet{Spiniello2015} 
and other two companion papers of the Survey, in preparation. Moreover,
we already presented results on the IMF slope of a single very massive lens galaxy used as pilot program to test our method in 
the XLENS I, \citet{Spiniello2011}.
In Section \ref{sec:ch1_overview} we introduce the XLENS Survey, highlighting its main goals and characteristics 
and we present its selection criteria.  In Section \ref{sec:ch1_observ} we describe   
the observations and data reduction process.
In Section \ref{sec:ch1_stellarkin} we present the spatially resolved kinematics up to $\sim 1$ \Reff\, obtained 
from VLT X-shooter spectra of 10 ETGs with $\sigmastar \geq250\,\mathrm{km\,s^{-1}}$.
We summarize our findings and present our preliminary conclusions in Section \ref{sec:ch1_summary}. 

\section{Overview of the XLENS Survey and target selection}
\label{sec:ch1_overview}

The XLENS aims to study 
the stellar population of strong early-type lens galaxies, for which exquisite HST multi-band images and 
very detailed lens models are available. 
The main goal of the survey is to disentangle the dark matter and stellar mass distributions, as well as to 
constrain the slope and the cut-off  mass of the low-mass end of the stellar IMF directly from spectra, 
by combining the lensing and dynamical results with the fully independent inferences from spectroscopic stellar population studies.

Our current sample has been selected from the Sloan Lens ACS Survey (SLACS, \citealt{Bolton2006, Bolton2008, Bolton2008b, Treu2006,Treu2009, 
Koopmans2006, Gavazzi2007, Gavazzi2008,  Auger2009, Auger2010b, Newton2011}). 
SLACS is a survey of lenses spectroscopically selected from the Sloan Digital Sky Survey (SDSS, \citealt{York2000}) 
with follow-up imaging conducted with the Advanced Camera and Spectrograph (ACS)  aboard the Hubble Space Telescope (HST). 
SLACS was initiated in 2003 with the purpose of confirming and imaging 
galaxies that act as strong gravitational lenses of emission-line background sources. 
The survey has greatly expanded the number of strong gravitational lens galaxies at redshift  
$z \leq 0.5$ with complete redshift information,  providing an ideal sample of systems 
to be analysed with joint lensing and dynamics techniques. 
In particular, half of the lenses were selected from the SDSS luminous red galaxy 
(LRG) spectroscopic sample (\citealt{Eisenstein2001}). 
The LRG sample is defined by photometric selection cuts that very 
efficiently select massive ETGs in the redshift range $0.15 < z \le 0.5$, as confirmed by
SDSS spectroscopy. These galaxies are very homogeneous in their
spectral, photometric, and morphological properties. 
The remaining targets were selected with the same spectroscopic selection procedure from
within the MAIN galaxy sample of SDSS (\citealt{Strauss2002}).
 The MAIN sample is more heterogeneous, and therefore
 a quiescent, absorption-dominated spectral criterion was imposed by requiring the lens candidates to 
 have rest-frame equivalent widths in H$\alpha$ of EW$_{\mathrm{H}\alpha} < 1.5$ \AA. 
 The method by which lens candidate were selected is extensivey described by \citet{Bolton2006}. 

The current SLACS sample consists of 131 strong gravitational lens 
candidates (\citealt{Bolton2008, Treu2009, Auger2009, Shu2015}).
To date, this represents the largest uniform sample of strong gravitational lenses and it includes
 98 ``grade-A'' and 33 ``grade-B'' and ``grade-C'' (likely lenses) confirmed strong galaxy-galaxy 
lens systems complete with lens and source redshifts, F814W lens-galaxy photometry, gravitational lens models,
and measured stellar velocity dispersions (\citealt{Auger2009}). 
Approximately 80\% of the ``grade-A'' systems have elliptical morphologies while
$\sim 10$\% show some spiral structure; the remaining lenses have lenticular morphologies.
 Such a large and high-quality lens sample provides a unique resource for the quantitative study of massive galaxies.

SLACS is a lensing-selected sample and hence also mass-selected sample, largely peaking around  $1-2$M$^*$ galaxies. 
\citet{Shu2015}, published after the XLENS data was acquired, increased the final sample extending it to lenses below below M$^*$ 

From the final SLACS sample of $\sim 100$ confirmed lenses, 
the XLENS sample includes only ``grade-A'' systems with elliptical 
morphology that lie above the knee of the luminosity function 
($\sigma_{\star, \rm SDSS}>250$ \kms)\footnote{SLACS is a lensing-selected sample and 
hence also mass-selected sample, largely peaking around  $1-2$M$^*$ galaxies}.

The velocity dispersion cut has been made because of the finding 
that there is a trend of increasing fraction of ``non-luminous" mass beyond M$^*$ galaxies 
(\citealt{Treu2010, Auger2009, Auger2010, Barnabe2013}).  
This monotonically increasing of the mass-to-light ratio as a function of  
the velocity dispersion could be due to an increasing number of low-mass stars 
 (i.e. a  low-mass IMF steeper than the Milky Way, \citealt{Kroupa2001, Chabrier2003}) 
 or to an increase in internal dark matter, or a combination thereof.

We note that this fraction 
is very low (less than 10\%) around M$^*$ (\citealt{Auger2009, Auger2010}). 
Although below M$^*$ this fraction might go up again, as suggested by other 
dynamical and gravitational lensing measurements, 
such low-mass lenses 
are rare and much harder to study (see \citealt{Shu2015}). 
The current XLENS sample does therefore not include any of these only recently discovered 
lower-mass lens galaxies. 

The current XLENS-SLACS sample consists of eleven 
luminous red galaxies  (including one previously published system not selected from SLACS) 
with apparent $V$-magnitudes between $15.8$ and $18.1$ (from \citealt{Auger2009}), high-resolution and high signal-to-noise (S/N) 
HST imaging in B, V, I and H-bands, and with detailed lensing models constructed by the SLACS collaboration. 

In addition to the lens galaxies selected from SLACS, 
another system has been observed in X-shooter Guaranteed Time (GTO, P087.A-0620):
SDSSJ1148+1930, also known as the ``Cosmic Horseshoe'' (\citealt{Belokurov2007}). 
This lens at redshift $z=0.444$ was analysed as a ``pilot program''
for the survey. Results on this extremely massive galaxy 
($\langle \sigmastar \rangle(\leq$\Reff) = $352\pm10\pm16\,\mathrm{km}\,\mathrm{s}^{-1}$) were presented by \citet{ Spiniello2011}. 
In the following sections, we limit our discussion to the ten SLACS-selected XLENS systems (with $0.11 \leq z \leq 0.44$). 
 
\begin{figure}
\includegraphics[height=6.5cm]{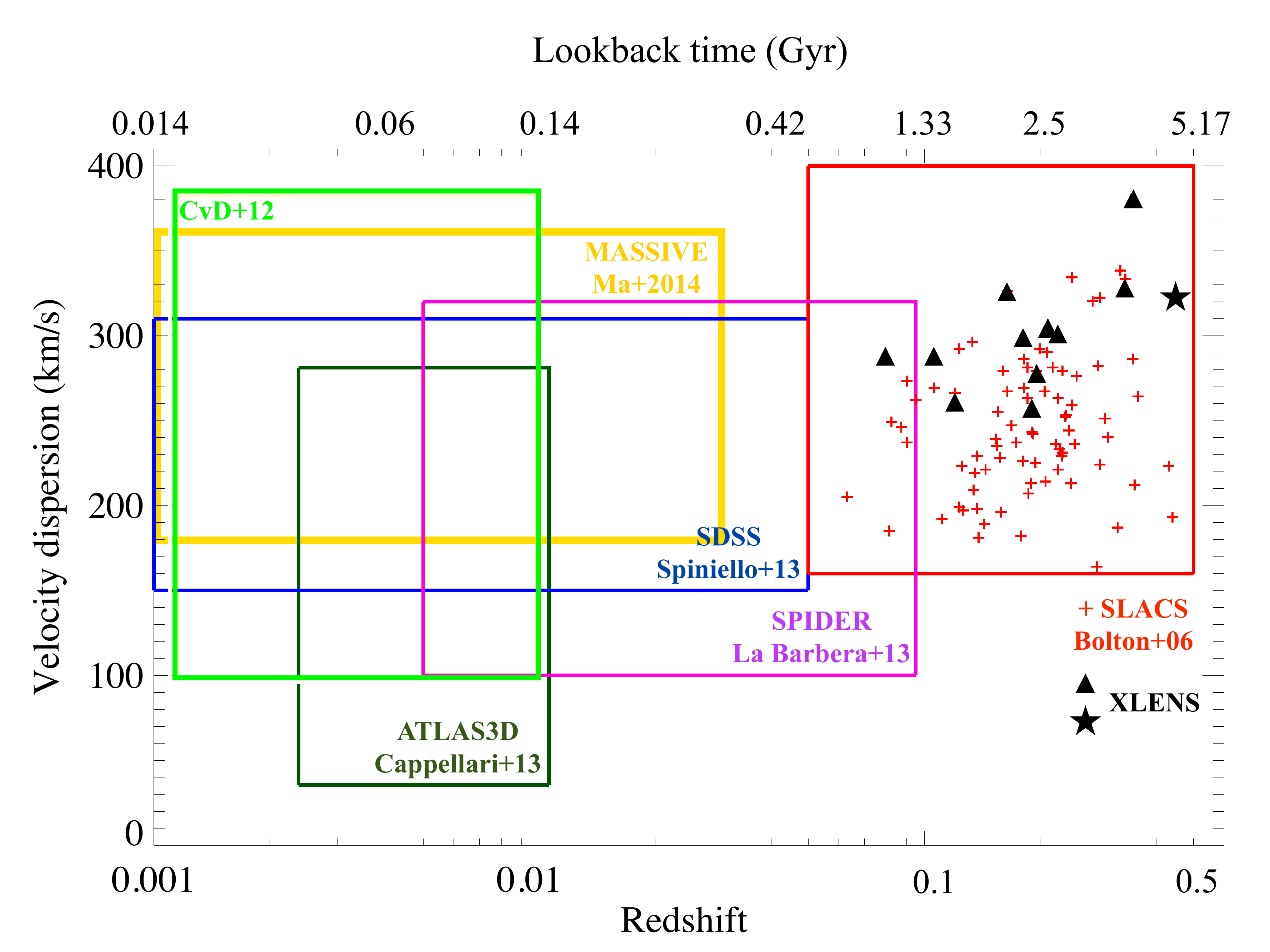}
\caption{Parameter space of the XLENS sample (XLENS from SLAC: black triangles, Pilot Program: black star) 
compared with other ETG samples (colored box).  The SLACS systems are plotted as red crosses. 
Some overlap at the high-mass exists between all these studies but the XLENS galaxies extend the redshift range until a lookback time of 5 Gyr.}
\label{fig:xlensintro_paramspace}
\end{figure}

In Figure~\ref{fig:xlensintro_paramspace} the mass-redshift space covered by the final XLENS sample (black triangles and star) 
is compared to those from the SLACS  parent sample (red crosses) as well as other sample of massive ETGs that have been
used to study the low-mass end of the IMF (\citealt{Cappellari2013, Conroy2012,LaBarbera2013, Ma2014}, colored boxes). 
The different samples overlap in the high-mass regime but XLENS pushes the redshift boundaries beyond the local Universe.

The current sample of lens ETGs with their photometric parameters and physical properties is presented in Table~1. 

\ctable[
caption = Properties of observed systems., 
star,
label = xlensintro_obsSLACS
]{lllllll}{
\tnote[1]{ This system is not selected from SLACS but it satisfies the LRG criteria and was observed in a similar set-up during the XLENS pilot program (\citealt{Spiniello2011}).} 
\tnote[2]{From the F606W filter on WFPC2.}
 
}{
\hline
\hline
{\bf SLACS System}  &z$_{\rm lens}$ & z$_{\rm BG}$ & \Reff (kpc) & R$_{\rm Ein}$ (kpc)  & m$_{\rm V}$ (mag)$^{2}$ & PA slit ($^{\circ}$)\\
\hline
SDSSJ0037-0942	&	0.1955	&	0.6322		&	7.03	&		4.95		&		16.90 		&		 11.4\\ 
SDSSJ0044+0113	&	0.1196	&	0.1965		&	5.56	&		1.72		&		16.32 		&		 151.3\\ 
SDSSJ0216-0813 	&	0.3317	&	0.5235		&	12.6	&		5.53		&		18.36  		&	81.2	\\ 
SDSSJ0912+0029 	&	0.1642	&	0.3239		&	10.8	&		4.58		&		16.56 		&	11.7	\\ 
SDSSJ0935-0003 	&	0.3475	&	0.4670		&	20.7	&		4.26		&		17.71 		&	145.2	\\ 
SDSSJ0936+0913 	&	0.1897	&	0.5880		&	6.61	&		3.45		&		17.12 		&	145.3	\\ 
SDSSJ0946+1006 	&	0.2219	&	0.6085		&	8.33	&		4.95		&		17.78 		&	10.3	\\ 
SDSSJ1143-0144 	&	0.1060	&	0.4019		&	9.21	&		3.27		&		15.83 		&	118.7	\\ 
SDSSJ1627-0053 	&	0.2076	&	0.5241		&	6.66	&		4.18		&		16.91 		&		6.9 \\ 
SDSSJ2343-0030 	 &	0.1810	&	0.4630		&	8.27	&		4.62		&		17.17 		&		144.0\\ 
SDSSJ1148+1930$^{1}$ & 0.4440 & 2.3815 & 12.5 & 29.0 &  20.02 & 99.0 \\
\hline
\hline
}

\section{Observations and data reduction}
\label{sec:ch1_observ}
X-shooter (\citealt{Vernet2011}) observations of the all systems were carried out in a total of $30$ hours of Observation 
Time (OT), awarded in two periods (P086 and P089)\footnote{$15$ hours OT during P86 (P086.A-0312, PI: Koopmans) and $15$ hours 
OT during P89 (P089.A-0364, PI: Spiniello).}. 
 The long-slit mode was used, splitting the beam over three arms: UVB (R=3300 with
a $1\farcs6$ slit); VIS (R=5400 with a $1\farcs5$ slit); and NIR (R=3300
with a $1\farcs5$ slit), covering the wavelength range from 3000 to 25000 \AA\ simultaneously.  
The 11\arcsec\ long slit was always centred on the lens
 galaxy with the position angle (PA) aligned with the major axis of the systems or with an angle
that minimizes the contamination from the lensed background source and leaves enough sky region
to facilitate accurate sky subtraction.  
A different number of Observation Blocks (OBs) have been obtained for each system
to reach a S/N high enough to perform stellar population analyses 
(S/N $\sim 50$ per \AA). 
Each OB had the same layout with three scientific exposures for a total 
exposure time on target for each arm of $\sim 2500$ s  per 
OB\footnote{In the NIR arm, the total exposure time is the product of DIT (Detector Integration Time), 
NDIT (Number of DITs) and NINT (Number of INTegrations): T = DIT $\times$ NDIT $\times$ NINT. 
We used for each of the 3 scientific exposures 3 DITs of 278 sec.}.  
During the observations, the seeing varied from $\sim0\farcs6$ to $\sim0\farcs9$.  
Standard calibration frames and standard stars for flux calibration
were obtained after each OB. \\
\indent The data have been reduced using the ESO X-shooter pipeline v. 2.0.0 
(\citealt{Goldoni2006}) and the Gasgano data file organiser developed by ESO. 
The pipeline reduction performs standard bias subtraction and flat-fielding of the
raw spectra. Cosmic rays are removed using LACosmic (\citealt{vanDokkum2001}).
For each arm, we extract the orders and rectify them in wavelength
space using a wavelength solution previously obtained from the
calibration frames.  The resulting rectified orders are then shifted and
co-added to obtain the final two-dimensional (2D) spectrum.  
We extract a one-dimensional spectrum (1D) from the resulting 2D merged spectrum,
using our own IDL code that 
also produces the corresponding error file and bad pixel map.  
The final signal-to-noise ratio in the UVB+VIS
spectrum varies from system to system, with a minimum of $\sim 50 $ and a 
maximum of  $\sim 145$ per Angstrom ($0.2$ \AA /${\rm pix}$). 
We finally perform an iterative sigma-clipping to
clean the spectrum of any residual bad pixels, sky lines and cosmic rays.
No telluric correction has been applied so that prominent atmospheric 
absorption bands can still be seen in
the final UVB+VIS spectra shown in rest-frame wavelengths in Figure~\ref{fig:xlensintro_spectra1}(a) 
and Figure~\ref{fig:xlensintro_spectra2}(b). 
For some of the systems, it is also possible to see emission lines from the background source, 
although the chosen PA minimizes their contribution to the stellar spectra. 
We mask these lines as well as telluric absorption when extracting the spatially-resolved kinematic profile.\\
\indent Because the near-infrared spectrum still suffers seriously from sky-line
residuals that are difficult to remove using the current pipeline, we limit ourselves in this paper to the
UVB-VIS region of the spectrum and defer a full analysis of the infrared data to future publications. 

 \begin{figure*}
\centering
\includegraphics[height=21cm]{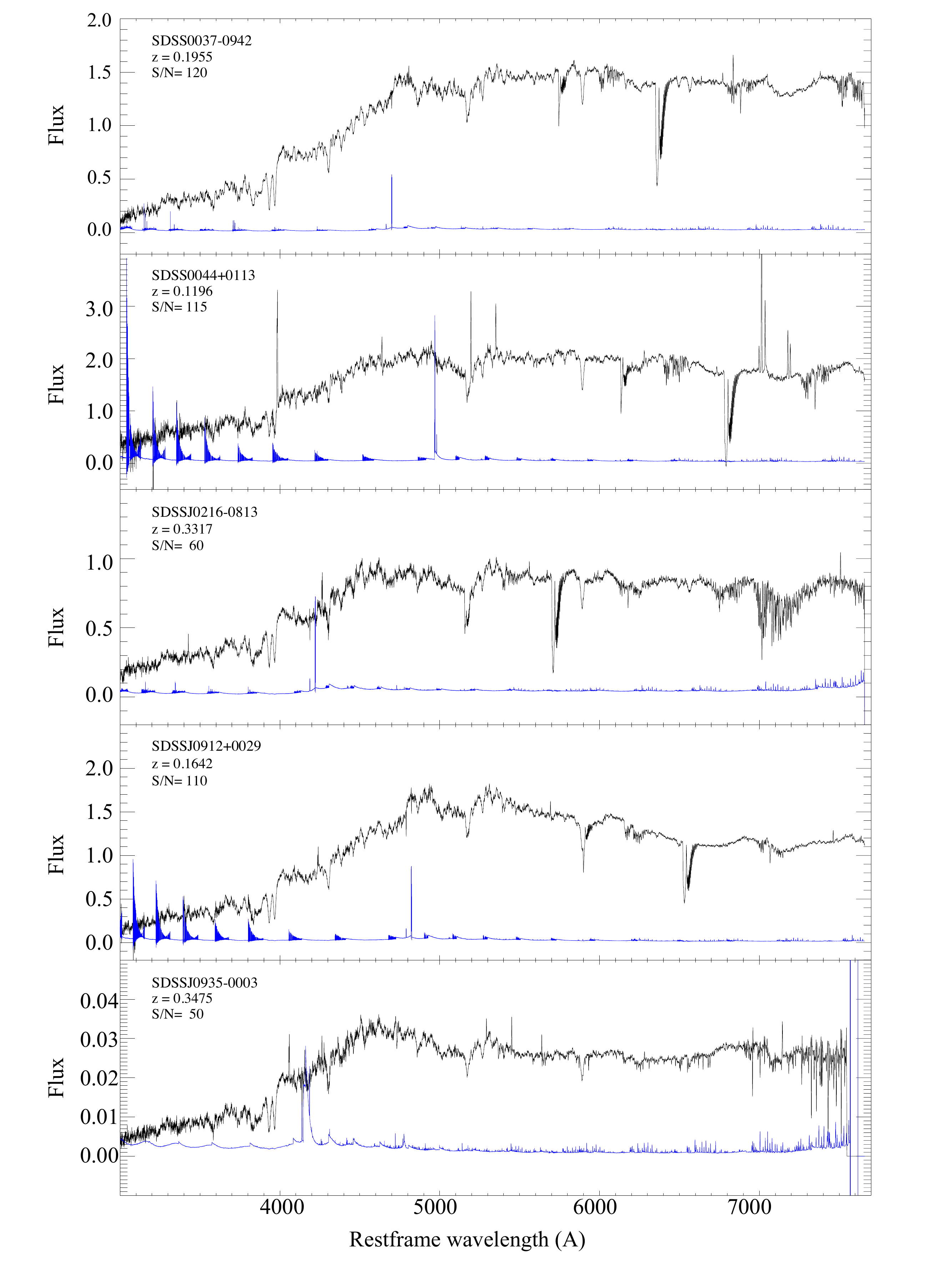}
\caption{ {\bf (a).} Final luminosity-weighted UVB--VIS 1D X-shooter spectra (in black) and respective errors (in blue)
 extracted from a rectangular aperture of $\sim 2\arcsec \times 1\farcs5$ 
 centered on the galaxies, shifted to the rest-frame and smoothed (3-pixel boxcar) for displaying purposes only. 
 IDs, redshifts of the lens and average S/N (per \AA) are shown in the panels.  Flux is in units of $10^{-16} \mathrm{erg/s/cm}^{2}/$\AA. 
 Telluric absorption lines and emission coming from the background source have not been removed from the spectra. 
 The peak in the error spectra between $4500$ and $5500$ \AA\  shows the point at which UVB and VIS 1D spectra have been joined. }
\label{fig:xlensintro_spectra1}
\end{figure*} 

 \begin{figure*}
\ContinuedFloat
\centering
\includegraphics[height=21cm]{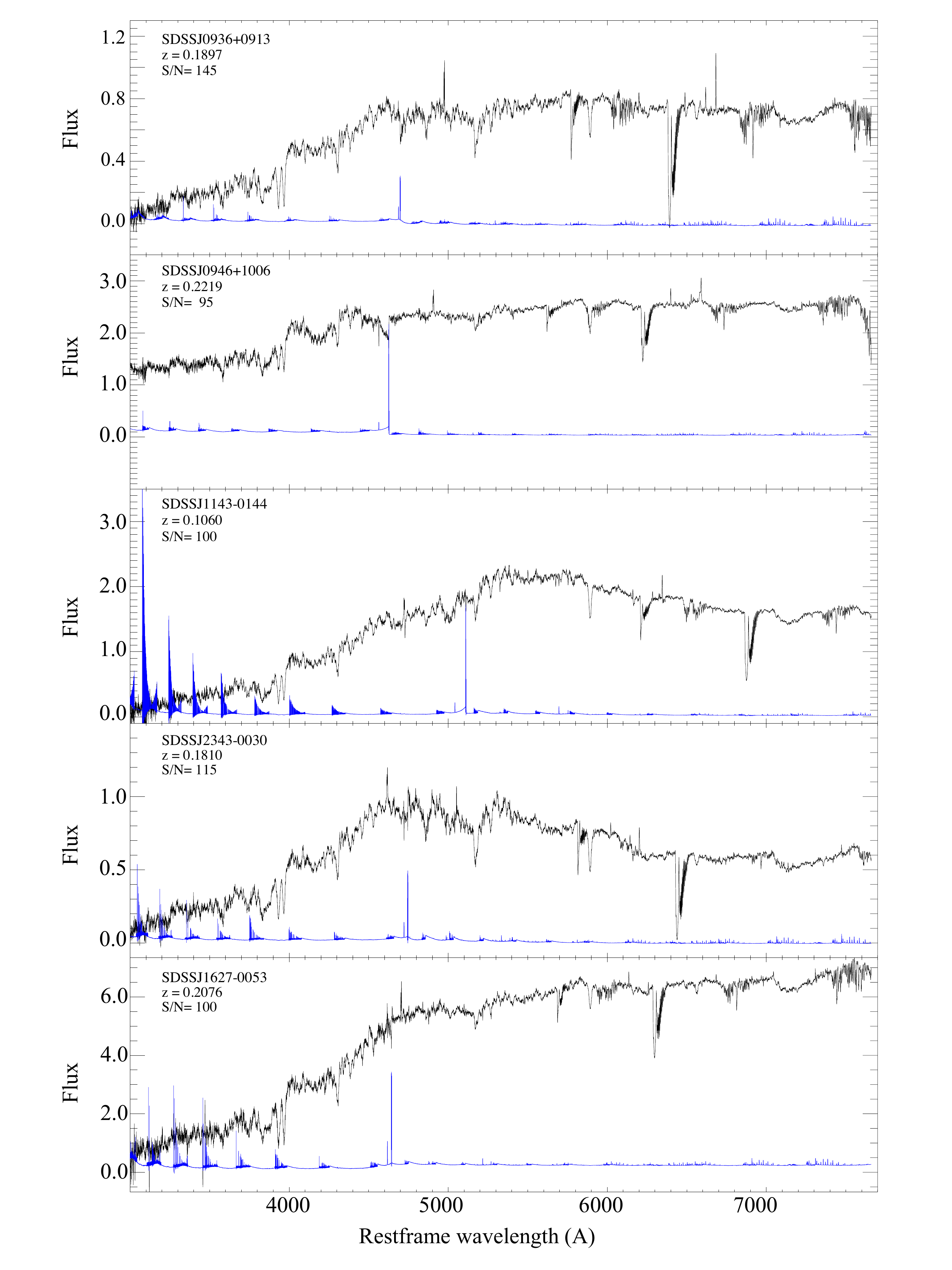}
\caption{ {\bf (b).} continued }
\label{fig:xlensintro_spectra2}
\end{figure*} 

\section{Stellar Kinematics}
\label{sec:ch1_stellarkin}

\subsection{Global Kinematics}
\begin{figure}
\includegraphics[height=11.5cm]{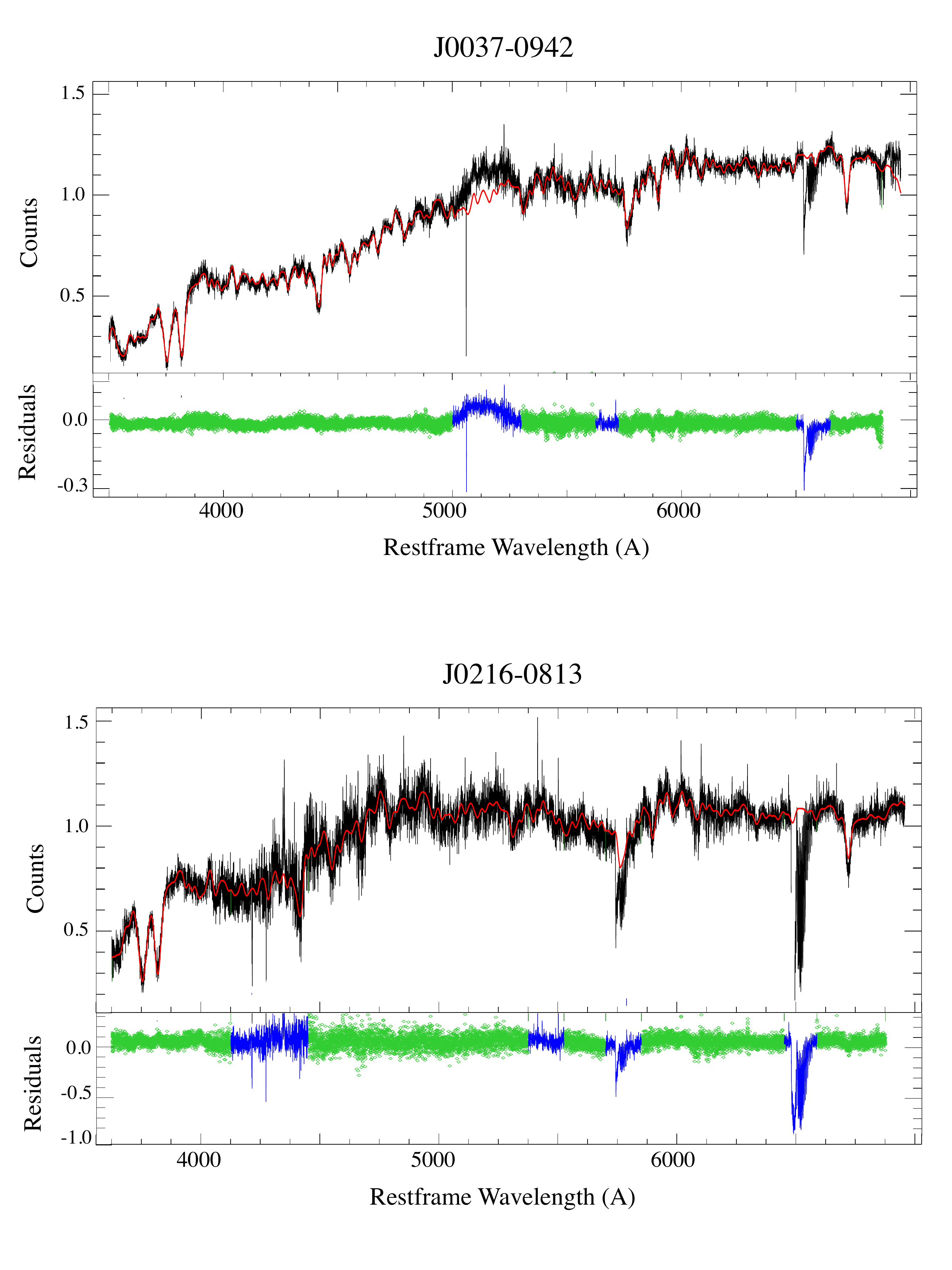}
\caption{Two examples of pPXF fits. The upper row shows one of the best cases ($\chi^{2}/{\rm DOF} \sim 1.15$), 
while the lower row shows the worst case ($\chi^{2}/{\rm DOF} \sim 1.65$). 
\textit{Top panels:} UVB+VIS galaxy
  spectrum (black) and correspondent best-fit template
  (red), both in restframe wavelength. \textit{Bottom panels:}
  Residuals from the fit. Bad-pixels excluded from the fitting
  procedure (sky line, telluric lines) are shown in blue. See the text
  for more information.}
\label{fig:xlensintro_pPXF_prof}
\end{figure}

We measure the luminosity-weighted line-of-sight velocity
dispersion (LOSVD) of the lens galaxies from the final 1D UVB--VIS spectra 
extracted from a rectangular aperture of $2\arcsec \times 1\farcs5$ 
centred on the galaxy. 
We use the Penalized Pixel Fitting (pPXF) code of \citet{Cappellari2004} to 
determine the combination of stellar templates
which, when convolved with an appropriate LOSVD, best reproduces the galaxy spectrum.  The best-fitting
parameters of the LOSVD are determined by minimizing a $\chi^{2}$
penalty function, yielding the mean velocity and 
velocity dispersion ($v$ and $\sigmastar$, respectively) and their uncertainties. 
In Figure~\ref{fig:xlensintro_pPXF_prof} we show examples of pPXF fits for the
 highest and lowest signal-to-noise systems. 

We focus on the absorption lines between 3500--6500 \AA\ (including Ca K
and H, G4300, H$_{\beta}$, Mg$_{b}$, strong Fe lines, NaD and some TiO molecular absorption bands). 
To minimize errors due to potential mismatches between the resolution of the templates and the galaxy spectrum, 
we use X-shooter spectra obtained as part of the
X-shooter Stellar Library (XSL) survey\footnote{http://xsl.u-strasbg.fr} (\citealt{Chen2014}),
with higher instrumental resolution. For the galaxy spectrum we use a 1.5-arcsecond slit (corresponding to
 $\langle \sigma_{\rm instr}\rangle \sim 35$ \kms, in the observed frame, and to 
$\langle \sigma_{\rm instr}\rangle \sim 28$ \kms$\,$ in the restframe) 
while for the stellar templates the slit width are $0\farcs5$ and $0\farcs7$ in UVB and VIS respectively,  
corresponding to $\langle \sigma_{\rm instr}\rangle \sim 12$ \kms.
As a test of the accuracy of our measurements, we use the more
heterogeneous MILES\footnote{http://www.iac.es/proyecto/miles/pages/stellar-libraries/miles-library.php}
stellar template library by \citet{SanchezBlazquez2006}.  We
select 100 stars (F, G, K, M) in the range 3525--7500 \AA, with
2.5 \AA\ FWHM spectral resolution (corresponding roughly to $30 \leq \sigma_{\rm MILES} \leq 70$ \kms 
$\,$ in the selected wavelength range), which is comparable to the XLENS spectra resolution 
in the selected regions ($\sigma_{\rm XSH} \sim 35$ \kms) 
that have been smoothed to the same spectral resolutions with a Gaussian. 
Fitting with different stellar templates does not appreciably affect the resulting LOSVD.
As a second test, we fit the same stellar templates to two different spectral regions for each galaxy separately 
(blue region: $3800-5000$~\AA\  and red region: $5000-6500$~\AA ). 
Only for two systems (SDSSJ$0936$+$0913$ and SDSSJ$1143$-$0144$) 
do we find slightly different results between the blue and the red parts of the spectrum, 
but the results are always consistent within $2\sigma$. 
The scatter between the different fits is used to estimate additional systematic uncertainties related to
template mismatch and spectral coverage. These errors are folded into the final error budget. 
We report our results and the comparison with other published velocity dispersion 
 results for each ETG in Table~\ref{tab:xlensintro_kinematics}.
We find a good agreement between our values extracted from a rectangular aperture of $2\arcsec \times 1\farcs5$ 
 and the SDSS velocity-dispersion measurements obtained in the $3\arcsec$-diameter spectroscopic 
fiber (for most of the systems within $1\sigma$ error). 
We also find good agreement with the results by \citet{Czoske2012} obtained from (a larger) aperture-integrated
spectra from the VIMOS Integral Field data on five systems that overlap with the XLENS sample. 

 \begin{table*}
 \center
  \caption{Luminosity-weighted stellar kinematics of the lenses. $\langle \sigma_{XSH} \rangle$ has been 
 extracted from a rectangular aperture of $2\arcsec \times 1\farcs5$ centered on the galaxies, while 
  $\sigma_{SDSS}$ is measured within an aperture with a diameter of $3\arcsec$. VIMOS velocity dispersions 
 were measured from aperture-integrated spectra from data from the VIMOS Integral-Field Unit (\citealt{Czoske2012}).}
\label{tab:xlensintro_kinematics}
 \begin{tabular}{lccc} 
\hline
\textbf{ID name}&\textbf{$\langle \sigma_{\rm XSH} \rangle$ (\kms)} &  \textbf{$\sigma_{\rm SDSS}$ (\kms)} &  \textbf{$\sigma_{\rm VIMOS}$ (\kms)} \\
\hline 
SDSSJ0037-0942    & 	$ 277 \pm 6$    &$279 \pm 14$ & $ 245.3^{+6.9}_{-7.2} $\\
SDSSJ0044+0113   &	 $260 \pm 8$    &$266 \pm 13$ 	& ...\\
SDSSJ0216-0813   &	  $ 327\pm 19$  &$333 \pm 23$	&  $ 340.7^{+7.8}_{-7.7}$\\
SDSSJ0912+0029   &	 $ 325 \pm 10$   &$326 \pm 16$	&  $ 306.5^{+10.9}_{-11.4}$\\
SDSSJ0935-0003   &	 $ 380 \pm 22$   &$396 \pm 35$	&  $ 330.4^{+9.0}_{-8.5}$\\
SDSSJ0936+0913  &	  $ 256 \pm 18 $  &$243 \pm 12$	& ...\\
SDSSJ0946+1006  &	  $ 300 \pm 22 $  &$263 \pm 21$ & ...	\\
SDSSJ1143-0144   &	  $287 \pm 18 $  &$269 \pm 10$	& ...\\
SDSSJ1627-0053   &	  $ 303 \pm 23$  &$290 \pm 20$	&  $ 272.6^{+7.8}_{-8.9}$\\
SDSSJ2343-0030 &   $ 298 \pm 21 $ &$269 \pm 16$ 	& ...\\
\hline
\end{tabular}
\end{table*}

\subsection{Spatially-Resolved Kinematics}
To extract the spatially-resolved kinematic information, we define a number of  
spatially-varying apertures (with adequate S/N ratio, $\mathrm{S/N}>10$) along the radial direction for each galaxy, 
and we sum the signal within each aperture.
Apertures are always defined to be larger than the seeing in order to have approximately independent 
kinematic measurements for each aperture.   
The streaming motion and velocity dispersion are measured from each
spatially-resolved spectrum using pPXF as described above. 
Also in this case we perform tests to check our error determinations: 
we fit the blue and red spectral regions separately, masking out the 
most prominent telluric lines in the VIS-red range; and we use different 
stellar templates from the two different stellar libraries 
(eleven XSL stars of G, K and M spectral types and a sub-selection of MILES stars of the same spectral types).  
The uncertainties on the inferred kinematics are estimated by adding 
in quadrature the formal uncertainty given by pPXF and the
scatter in the results for different templates and spectral regions.

Our goal is to determine the kinematic profiles up to about one effective radius for all the systems. 
The first and second moments of the LOSVD out to their effective radii, complemented
with two-integral axisymmetric models (\citealt{Barnabe2007, Barnabe2009})
allow us to obtain a precise measurement of the logarithmic slope of the total mass density profile of
the lens galaxy (e.g.\ \citealt{Czoske2008, Barnabe2011,Barnabe2013, Spiniello2011}). 

\begin{figure*}
\begin{flushleft}
\centering
\includegraphics[height=13cm]{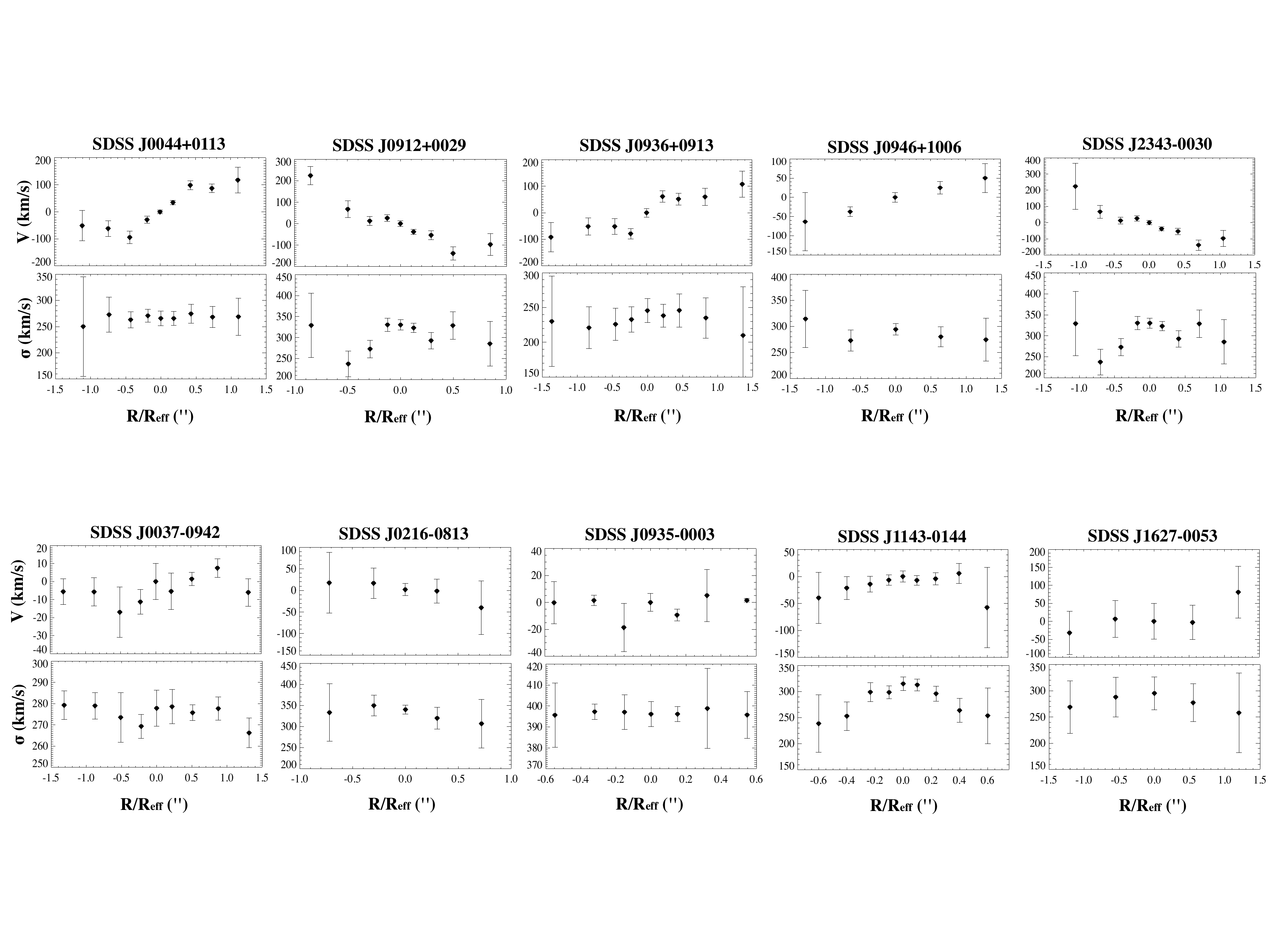}
\caption{Spatially-resolved kinematics of the lenses. Upper panels show the systems which have 
mild rotation and are classified as fast-rotators from Figure~\ref{fig:xlens_kinem}, lower panels show systems which do not  
rotate and are classified as slow-rotators. }
\label{fig:xlensintro_rotmap}
\end{flushleft}
\end{figure*}

The rotation and the velocity dispersion profiles for the ten XLENS-SLACS galaxies  
are shown in Figure~\ref{fig:xlensintro_rotmap}. 
The upper line shows systems with moderate rotation, whereas
 the lower line shows systems with almost-flat velocity profiles.
The weighted average values are always consistent within the formal 
error with the luminosity weighted values for an aperture of $2\arcsec \times 1\farcs5$.  

\begin{figure*}
\begin{flushleft}
\centering
\includegraphics[height=8cm]{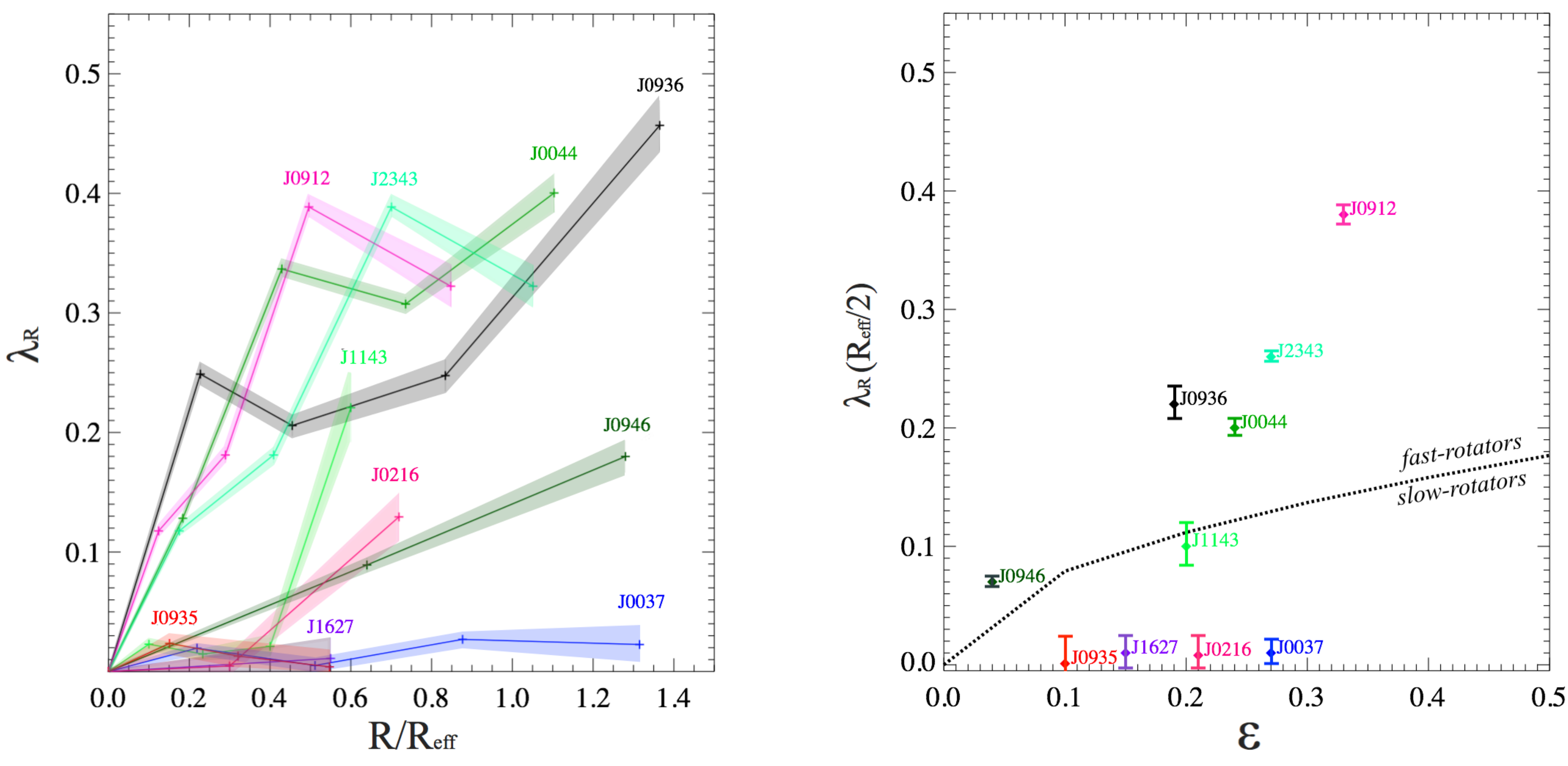}
\caption{{\sl Left:} Radial $\lambda_{R}$ profiles. The 1-sigma contours are showed with colored shaded regions.
{\sl Right:} The $\lambda_{R_{\rm eff}/2} - \epsilon$ relation for the XLENS galaxies. 
The dotted black line shows the division between slow- (below the line) and fast (above the line) rotators within $R_{\mathrm{eff}/2}$.
All most massive galaxies in our sample, except for J0912, belong to the slow-rotator group. }
\label{fig:xlens_kinem}
\end{flushleft}
\end{figure*}

We infer the averaged angular momentum ($\lambda_{R}$) using the definition of \citet{Emsellem2011}, 
with the spatially-resolved rotation velocity and velocity dispersion.  
We derive the radial $\lambda_{R}$ profiles (left panel of Figure~\ref{fig:xlens_kinem}) 
and plot them as a function of ellipticity ($\epsilon$), 
calculated from the lensing models ($\varepsilon = 1-q^*$, where $q^*$ is the axis ratio). 
This allows us to split the galaxy sample 
into slow- and fast-rotators. 
 The $\lambda_{R}$ parameter quantifies the (apparent) global
dynamical state of a galaxy and has been used by the ATLAS$^{\rm 3D}$ team (\citealt{Cappellari2011}) to show
that fast and slow rotators are physically distinct classes of galaxies with different orbital distribution and 
mass assembly histories. 
Simulations have demonstrated that galaxy mergers (major and minor) 
have a significant influence on the rotation properties: major mergers produce a significant increase in the 
angular momentum of the dark matter haloes at large distances from the center (\citealt{Vitvitska2002})
but it leads to a redistribution of the angular
momentum of the central stellar component outwards (e.g. \citealt{Bournaud2004}).
Fast rotators have either preserved or regained
their specific angular momentum in the central part and therefore 
they cannot have experienced major dry mergers which would
expel most of the angular momentum outwards and avoid the building 
of a rotation-dominated disc-like structure (\citealt{Emsellem2007, Naab2014}).
Concluding, the $\lambda_{R}$, interpreted as a proxy of the specific angular momentum, 
is an important quantity that allows one to link present-day kinematical profiles to the 
cosmological formation history of a galaxy.

The right panel of Figure \ref{fig:xlens_kinem} clearly shows that five of our galaxies are 
classified as slow rotators, whereas the other five are classified as fast rotators, but none of them show a very high  
$\lambda_{R}$ value. 
It also appears that the most massive galaxies in our sample tend to rotate more slowly than the least massive ones
(with the exception of J0912, which is classified as fast-rotator even though it has $\sigma$ $>$ $300$ $\kms$).
This result confirms the ATLAS$^{\rm 3D}$ result  that slow rotators dominate the high-mass end of ETGs. 

\section{Summary \& Conclusions}
\label{sec:ch1_summary}

In this paper, we have presented our spectroscopic follow-up survey of massive early-type lens galaxies: 
the X-shooter Lens Survey (XLENS). 
In particular, we 
\begin{itemize} 
\item highlight the scientific motivations and main goals of XLENS. 
\item present the current sample of eleven massive 
early-type lens galaxies, discussed the selection criteria and described the data reduction 
process for the XLENS-SLACS galaxies.
The remaining lens, also known as ``The Cosmic Horseshoe'', has been used as a pilot program of 
the survey and is the subject of the first paper of the series (XLENS I, \citealt{Spiniello2011}.
\item present the spatially resolved kinematic profiles up to $\simeq 0.8 - 1.5$ \Reff$\,$ for the lens galaxies. 
We found luminosity-weighted velocity dispersions mostly within $1\sigma$ agreement 
of those calculated from SDSS. Our velocity dispersions also agree with the results of \citet{Czoske2012} 
extracted from VIMOS integral-field spectra on 5 systems in common. 
\item derive the  $\lambda_{R}$ spatial profiles and calculate  $\lambda_{R_{\rm eff}/2}$ to split the 
sample into fast- and slow- rotators. We confirmed the ATLAS$^{\rm 3D}$ results that slow rotators 
dominate the high-mass end of the galaxy mass function.
 \end{itemize}

The combination of high-S/N spectroscopy from X-shooter with a strong gravitational-lensing
mass determination enables us to conduct an in-depth study of the dark and luminous matter 
within $\sim 1$\Reff \, and for the first time constrain the normalization, 
shape and cut-off mass (\Mcutoff) of the low-mass end of the IMF. 
We make use of the state-of-the-art joint lensing and kinematic code (\citealt{Barnabe2012}) 
and new SSP models (\citealt{Conroy2012}) that have been constructed specifically 
for the purpose of measuring the IMF slope down to $\sim 0.1\, \Msun$ 
for old, metal-rich stellar populations. 
The version of the SSP models that we use have been 
extended beyond their original parameter spaces, to explore super-solar [$\alpha$/Fe] and super-solar metallicity
and to disentangle elemental enhancements, metallicity changes and IMF variations in massive early-type galaxies (ETGs) 
with star formation histories different from the Solar neighborhood (following the method presented in \citealt{Spiniello2015a}).

In two following papers of the series, we use the joint lensing and dynamics code CAULDRON\footnote{`Combined 
Algorithm for Unified Lensing and Dynamics ReconstructiON', (\citealt{Barnabe2007,Barnabe2009, Barnabe2012})} 
to perform the analysis of the total mass distributions for the XLENS systems. 
We put constraints on the internal dark matter fractions, the orbital structure, the shape and flattening of the dark matter halo 
and investigate possible correlations of these quantities with physical parameters of the lens systems such as size, stellar density and stellar velocity dispersion.
Thanks to the high quality of our lensing and kinematics data, this technique  also makes it possible to put firm constraints on the stellar 
mass of the analyzed galaxies independently of assumptions 
about their stellar initial mass functions (IMF) or knowledge of their stellar populations. 

Moreover, we apply to the sample the line-index-based stellar population analysis presented in \citet{Spiniello2014a}
to study in detail the stellar contents of the systems and to constrain their integrated ages, metal abundances and
IMF slopes and to search for possible correlations between stellar population parameters and structural properties of the systems 
(i.e., mass, size, stellar density and/or velocity dispersion).

Finally, in a Letter already submitted, we study the impact of a non-universal IMF slope and/or a non-universal IMF low-mass 
cut-off on the inferred mass-to-light ratios by comparing the two completely independent constraints on the stellar mass, 
one from the fully self-consistent joint lensing+dynamics (L\&D) analysis, and the other from spectroscopy and SSP modelling. 
For this study we follow the approach presented in \citet{Barnabe2013} to obtain the joint 2D posterior PDF in the slope-\Mcutoff \, space. 

In the future, we plan to extend the current sample, observing lenses within the same range of velocity dispersions 
($\sigma_{\star, \rm SDSS}>250$ \kms) at higher redshift (up to $z\sim 1$) to test possible evolution of the IMF with 
cosmic time. We also plan to observe low-mass ETGs to build a complementary
dataset to our more massive ETGs with the purpose of covering a wider stellar  mass range and to test possible variation of the IMF slope with 
structural and stellar parameters, other than the stellar mass of the galaxy (e.g. total dynamical density, as we recently suggested in Spiniello et al. 2015).

\section*{Acknowledgments}
The authors thank the referee for providing constructive comments 
and help in improving the contents of this paper.\\
The use of the Penalized Pixel Fitting developed by Cappellari \&
Emsellem is gratefully acknowledged. Data were reduced using EsoRex and XSH
pipeline by ESO Data Flow System Group. 
M.B. acknowledges support from the Danish National Research Foundation. 
L.V.E.K. is supported in part through an NWO-VICI career grant (project number 639.043.308). 
T.T. acknowledges support from a Packard Research Fellowship.

\bibliographystyle{mn2e_fix}
\bibliography{XLENS_biblio.bib}

\label{lastpage}

\end{document}